\documentclass[
longbibliography
aps,
prl,
reprint,
superscriptaddress,
amsmath,
amssymb]{revtex4-2}
\usepackage{graphicx}
\usepackage{dcolumn}
\usepackage{bm}
\usepackage{gensymb}
\usepackage{comment}
\usepackage{amsmath}
\usepackage[colorlinks=true, citecolor=blue, urlcolor=blue, linkcolor=blue]{hyperref}

\begin{document}

\preprint{APS/123-QED}

\title{Eliminating Surface Oxides of Superconducting Circuits with Noble Metal Encapsulation}

\author{Ray D. Chang}
 \altaffiliation{These authors contributed equally to this work.}
\affiliation{Department of Electrical and Computer Engineering,
Princeton University, Princeton, New Jersey 08540, USA}

\author{Nana Shumiya}
 \altaffiliation{These authors contributed equally to this work.}
\affiliation{Department of Electrical and Computer Engineering,
Princeton University, Princeton, New Jersey 08540, USA}

\author{Russell A. McLellan}
\altaffiliation{Present address: Archer Aviation, San Jose, California 95134, USA}
\affiliation{Department of Electrical and Computer Engineering,
Princeton University, Princeton, New Jersey 08540, USA}

\author{Yifan Zhang}
\affiliation{Department of Electrical and Computer Engineering,
Princeton University, Princeton, New Jersey 08540, USA}

\author{Matthew P. Bland}
\affiliation{Department of Electrical and Computer Engineering,
Princeton University, Princeton, New Jersey 08540, USA}

\author{Faranak Bahrami}
\affiliation{Department of Electrical and Computer Engineering,
Princeton University, Princeton, New Jersey 08540, USA}

\author{Junsik Mun}
\affiliation{Center for Functional Nanomaterials, Brookhaven National Laboratory, Upton, New York 11973, USA}
\affiliation{Condensed Matter Physics and Materials Science Department, Brookhaven National Laboratory, Upton, New York 11973, USA}

\author{Chenyu Zhou}
\affiliation{Center for Functional Nanomaterials, Brookhaven National Laboratory, Upton, New York 11973, USA}

\author{Kim Kisslinger}
\affiliation{Center for Functional Nanomaterials, Brookhaven National Laboratory, Upton, New York 11973, USA}

\author{Guangming Cheng}
\affiliation{Princeton Materials Institute, Princeton University, Princeton, New Jersey 08540, USA}

\author{Alexander C. Pakpour-Tabrizi}
\affiliation{Department of Electrical and Computer Engineering,
Princeton University, Princeton, New Jersey 08540, USA}

\author{Nan Yao}
\affiliation{Princeton Materials Institute, Princeton University, Princeton, New Jersey 08540, USA}

\author{Yimei Zhu}
\affiliation{Condensed Matter Physics and Materials Science Department, Brookhaven National Laboratory, Upton, New York 11973, USA}

\author{Mingzhao Liu}
\affiliation{Center for Functional Nanomaterials, Brookhaven National Laboratory, Upton, New York 11973, USA}

\author{Robert J. Cava}
\affiliation{Department of Chemistry,
Princeton University, Princeton, New Jersey 08540, USA}

\author{Sarang Gopalakrishnan}
\affiliation{Department of Electrical and Computer Engineering,
Princeton University, Princeton, New Jersey 08540, USA}

\author{Andrew A. Houck}
\affiliation{Department of Electrical and Computer Engineering,
Princeton University, Princeton, New Jersey 08540, USA}

\author{Nathalie P. de Leon}
\email{npdeleon@princeton.edu}
\affiliation{Department of Electrical and Computer Engineering,
Princeton University, Princeton, New Jersey 08540, USA}

\date{\today}

\begin{abstract}

The lifetime of superconducting qubits is limited by dielectric loss, and a major source of dielectric loss is the native oxide present at the surface of the superconducting metal. Specifically, tantalum-based superconducting qubits have been demonstrated with record lifetimes, but a major source of loss is the presence of two-level systems (TLSs) in the surface tantalum oxide. Here, we demonstrate a strategy for avoiding oxide formation by encapsulating the tantalum with noble metals that do not form native oxide. By depositing a few nanometers of Au or AuPd alloy before breaking vacuum, we completely suppress tantalum oxide formation. Microwave loss measurements of superconducting resonators reveal that the noble metal is proximitized, with a superconducting gap over 80\% of the bare tantalum at thicknesses where the oxide is fully suppressed. We find that losses in resonators fabricated by subtractive etching are dominated by oxides on the sidewalls, suggesting total surface encapsulation by additive fabrication as a promising strategy for eliminating surface oxide TLS loss in superconducting qubits.

\end{abstract}

\maketitle


Superconducting qubits are the basis of many large scale quantum processors, enabling demonstrations of quantum error correction \cite{sivak_real-time_2022, acharya_suppressing_2022,dualrail2024}, quantum many body physics \cite{Kollár2019, mi_time-crystalline_2022, andersen_observation_2022, andersen2024}, and quantum simulation \cite{karamlou2022, jgcm2023, karamlou2024}.  Despite this progress, single qubit coherence remains a major limiting factor in building scalable processors based on superconducting qubits. Single qubit coherence is limited by dielectric loss, particularly at surfaces and interfaces \cite{read_precision_2022, de_leon_materials_2021, krupka_complex_1999, krupka_use_1999}. Tantalum-based superconducting qubits have recently been discovered to enable record lifetimes and coherence times \cite{placeRodgers2021,wang2022,ganjam2024}. Losses in state-of-the-art tantalum devices are dominated by TLSs in surface oxides and the bulk substrate \cite{crowley2023,mclellan_xps2023}. Avoiding the formation of surface oxides would eliminate this loss channel, and recent studies have demonstrated that encapsulation with other materials can mitigate losses associated with surface oxides of niobium \cite{bal2023, deory2024}, pointing to a potential strategy for avoiding surface losses in tantalum (Ta).

Here we encapsulate Ta superconducting resonators by depositing a noble metal (gold or gold-palladium) after the Ta deposition, before breaking vacuum. We characterize the film using x-ray photoelectron spectroscopy (XPS), scanning transmission electron microscopy (STEM), electron energy loss spectroscopy (EELS), and energy-dispersive x-ray spectroscopy (EDX) to show that the tantalum oxide is completely suppressed with only a few nanometers of noble metal encapsulation. The noble metal encapsulation layer is a normal (non-superconducting) metal, so it may contribute to additional loss if the normal metal is not fully proximitized by the underlying superconducting film \cite{belzig96,barends_minigap,gurevich17,Ustavshchikov2019}. We fabricate resonators with these encapsulated films, and perform microwave measurements to confirm the proximitization of the noble metal layer and to determine the effective superconducting gap of the heterostructure, which varies with encapsulation thicknesses. By fitting the dependence of the effective gap on encapsulation thickness with numerical solutions of the Usadel equations \cite{usadel_1970}, we estimate an interface transparency and find that it is consistent with a high quality, Ohmic contact. By comparing materials characterization with microwave device measurements, we find a wide range of encapsulation thicknesses that suppress the surface oxide while maintaining a high quality factor at base temperatures, from 3 nm to 26 nm. These results point to a promising strategy for eliminating surface oxide loss via total encapsulation and additive nanofabrication.

\begin{figure}[htbp]
\includegraphics[width=\linewidth]{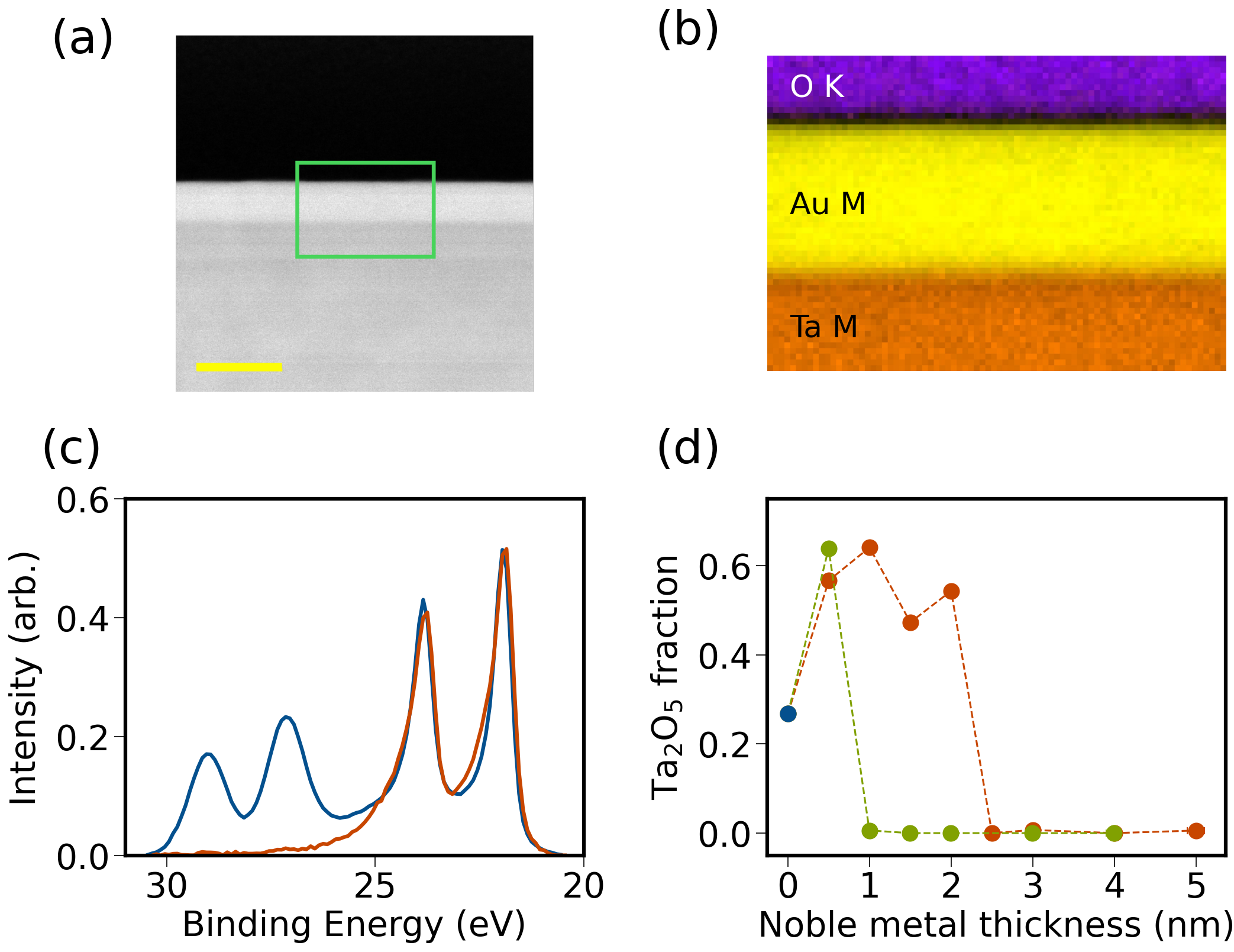}
\caption{\label{fig:fig1} Oxide suppression via noble metal encapsulation. (a) Survey annular dark-field (ADF) image of encapsulated Ta film cross section, green box indicates cross-sectional STEM region in (b). The scale bar is 20 nm long. (b) False colored EELS elemental map overlaid over ADF image shows spatial distribution of Ta (orange), Au (yellow), and oxygen (purple), found by integrating over the respective ionization edge. The Ta-Au interface is oxygen-free, demonstrating suppression of the tantalum oxide. The oxygen signal seen above the film comes from a permanent marker layer applied for protection against focused ion beam damage. (c) Ta4f XPS spectra of bare Ta (blue) and Ta encapsulated with Au (orange). The bare Ta spectrum exhibits four peaks, two from Ta metal and two from Ta$_2$O$_5$. The encapsulated Ta spectrum does not contain oxide peaks. (d) Atomic percentage of Ta$_2$O$_5$ for varying thicknesses of Au (orange) and AuPd (green) encapsulation, compared to the native oxide on a bare Ta film (blue). The Ta$_2$O$_5$ fraction is extracted from the ratio of normalized intensities of the Ta$_2$O$_5$ and metal Ta peaks. 2.5 nm of Au or 1 nm of AuPd is enough to suppress the tantalum oxide from forming. The vertical error bars reflect fit uncertainties and the horizontal error bars reflect instrumental thickness measurement uncertainties; for most data points these error bars are smaller than the plot markers}
\end{figure}

We deposit gold (Au) or gold palladium alloy (AuPd) on top of 200 nm thick $\alpha$-Ta films grown on 550 $\mu$m thick c-plane sapphire substrates with dc magnetron sputtering \cite{our_SI, Face_Prober_1987, Gladczuk2004, nakagomi2012,mukherjee2012}. The Au (AuPd) encapsulation layer is deposited \textit{in situ} right after the Ta film growth inside an ultra high vacuum chamber (with base pressure value of $5\times 10^{-9}$ Torr), avoiding the growth of native oxide, as verified by STEM [Figs. \ref{fig:fig1}(a,b)].

In order to quantify the degree of surface oxidation, we use XPS to measure the atomic percentage of tantalum pentoxide (Ta$_2$O$_5$) in the Ta4f spectrum [Figs. \ref{fig:fig1}(c,d)]. The Ta4f spectrum consists of 2 sets of doublet peaks, metallic Ta at 22 eV and 24 eV and Ta$_2$O$_5$ at 27 eV and 29 eV \cite{moulder1992, mclellan_xps2023} [Fig. \ref{fig:fig1}(c)]. The Ta$_2$O$_5$ peaks are not present in the binding energy spectrum of Au-encapsulated Ta, demonstrating that the amorphous surface oxides of Ta have been suppressed. 

We systematically investigate the encapsulation conditions needed to fully suppress the tantalum oxide by measuring Ta4f XPS spectra for various thicknesses of Au or AuPd encapsulation layers and extracting the intensity of the Ta$_2$O$_5$ peaks. Note that for very thin encapsulation layers, we report an effective thickness based on deposition rates, see \cite{our_SI} for more details. We obtain higher Ta$_2$O$_5$ intensity with nominal Au (AuPd) thickness less than 2.5 (1) nm, which we hypothesize is caused by an oxidation catalysis effect from small Au (AuPd) islands \cite{our_SI,hu2021,yang2022}. However, we find that 2.5 (1) nm of encapsulating Au (AuPd) is enough to fully suppress the tantalum oxide [Fig. \ref{fig:fig1}(d)], which indicates a material-dependent critical thickness at which the encapsulation film completely covers the Ta surface and forms an oxygen diffusion barrier.

We measure the degree of proximitization of the noble metal encapsulation by fabricating coplanar waveguide (CPW) quarter-wave resonators on various encapsulated Ta films with designs that were used in our prior work \cite{crowley2023}. The losses associated with TLSs and quasiparticles (QPs) can be characterized by measuring the resonance frequency as a function of temperature, or alternatively by measuring the internal quality factor as a function of intracavity photon number and temperature \cite{crowley2023,gao_jiansong_physics_2008}. From these measurements, the effective superconducting gap energy can be extracted \cite{crowley2023,our_SI}.

\begin{figure}[htbp]
\includegraphics[width=\linewidth]{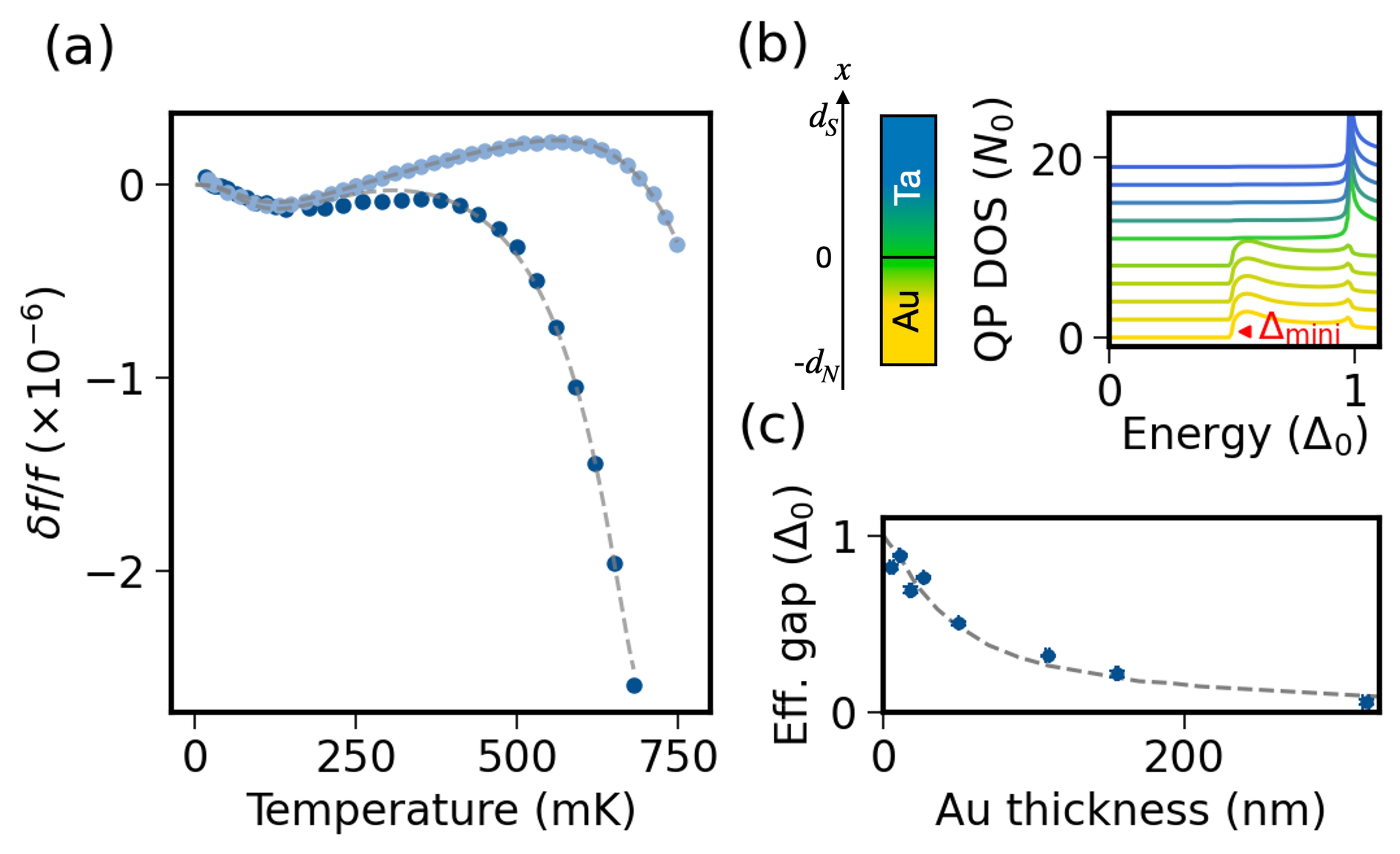}
\caption{\label{fig:fig2} Measuring the effective superconducting gap energy of encapsulated Ta films. (a) The measured frequency shift of two resonators with 50 nm (dark blue) and 5 nm Au (light blue) encapsulation layers as a function of temperature. The dashed grey lines are a fit to a model incorporating TLSs and quasiparticles \cite{crowley2023,our_SI}. Vertical error bars reflect fit uncertainties, and are smaller than the plot markers. (b) Cartoon of a Ta-Au superconductor-normal metal bilayer with respective Ta and Au thicknesses of $d_S$ and $d_N$ (left). Quasiparticle DOS in units of $N_0$, the normal-state DOS at the Fermi surface, at different positions in the Ta-Au bilayer (right). The density of states is obtained by numerically solving the Usadel equations. Different colors correspond to different locations in the bilayer represented by the color map on the left, and each curve is manually offset for ease of viewing. The red arrow denotes the location of the minigap's opening in the Au (normal metal) layer. (c) Effective gap in units of $\Delta_0$, the bulk Ta gap, extracted from microwave measurements of resonators with varying encapsulation thicknesses. The vertical error bars reflect uncertainties from the fit to the superconducting energy gap and the horizontal error bars reflect instrumental thickness measurement uncertainties. The dashed line is a fit to numerically calculated minigaps.}
\end{figure}

As a representative example, the frequency shift of resonators with 5 nm and 50 nm Au encapsulation layers that are otherwise identical is markedly different in the high temperature regime [Fig. \ref{fig:fig2}(a)]. In the low temperature regime, the non-monotonic dependence of frequency on temperature is dominated by TLSs \cite{quintana_characterization_2014,crowley2023}, and the two resonators exhibit similar behavior. At higher temperatures, the more rapid change in frequency with temperature is caused by thermal quasiparticles \cite{gao_jiansong_physics_2008, grunhaupt_2018}. With a thicker encapsulating layer we observe a larger change in the frequency shift at high temperature, indicating a larger population of thermal quasiparticles [Fig. \ref{fig:fig2}(a)]. These effects can be parameterized by a reduction in the effective superconducting gap energy of the film. This softening of the gap can be understood through the superconducting proximity effect \cite{belzig96,barends_minigap,gurevich17,Ustavshchikov2019}.

We quantitatively investigate the superconducting nature of encapsulated Ta films by extracting the effective gap of films undergoing additional iterative sputtering of Au to vary the encapsulation layer thickness from 5 to 323 nm [Fig. \ref{fig:fig2}(c)].  We observe a decrease in the effective gap in our devices as the encapsulation thickness increases. These results can be explained by distortions in the quasiparticle density of states (DOS) across the superconductor-normal metal interface caused by the proximity effect. Our measurements are particularly sensitive to higher populations of thermal quasiparticles induced by the opening of a superconducting energy gap smaller than the bulk Ta gap ($\Delta_0$) in the normal metal layer denoted as the minigap, $\Delta_{\text{mini}}$ [Fig. \ref{fig:fig2}(b)]\cite{belzig96,barends_minigap}. Here, the microwave resonator measurements are dominated by changes in the surface impedance of the film, and are thus most sensitive to the properties of the normal metal layer at the surface.

To understand the evolution of minigap size with increasing thicknesses of encapsulating Au, we fit our data to numerically calculated minigaps using the Usadel equations [Fig. \ref{fig:fig2}(c)]\cite{martinis_2000, degennes_1964, usadel_1970, our_SI, belzig98, ashcroft_mermin, bobrov2015point}. The only free fit parameter is the electron interface transparency, which we fit to a value of $t=0.24^{+.17}_{-.08}$ (see \cite{our_SI} for details on error estimations). This high interface transparency is consistent with the suppression of tantalum oxide and the formation of an Ohmic contact, and it is comparable with other reported transparency values of superconductor-normal metal interfaces grown via \textit{in situ} sputtering \cite{barends_minigap,hennings-yeomans2020,martinis_2000}.

For Au encapsulation thicknesses of around 3 nm where the tantalum oxide is completely suppressed [Fig. \ref{fig:fig1}(d)], the effective gap energy is still greater than 80\% of the bulk Ta gap. We observe an ideal range of encapsulation thicknesses from 3 to 26 nm over which encapsulated resonators maintain a high quality factor ($\approx10^7$) at base temperatures and high powers. These results indicate that encapsulated Ta films are a viable material platform for fabricating superconducting circuits with minimal reduction in the effective gap.

\begin{figure}[ht]
\includegraphics[width=\linewidth]{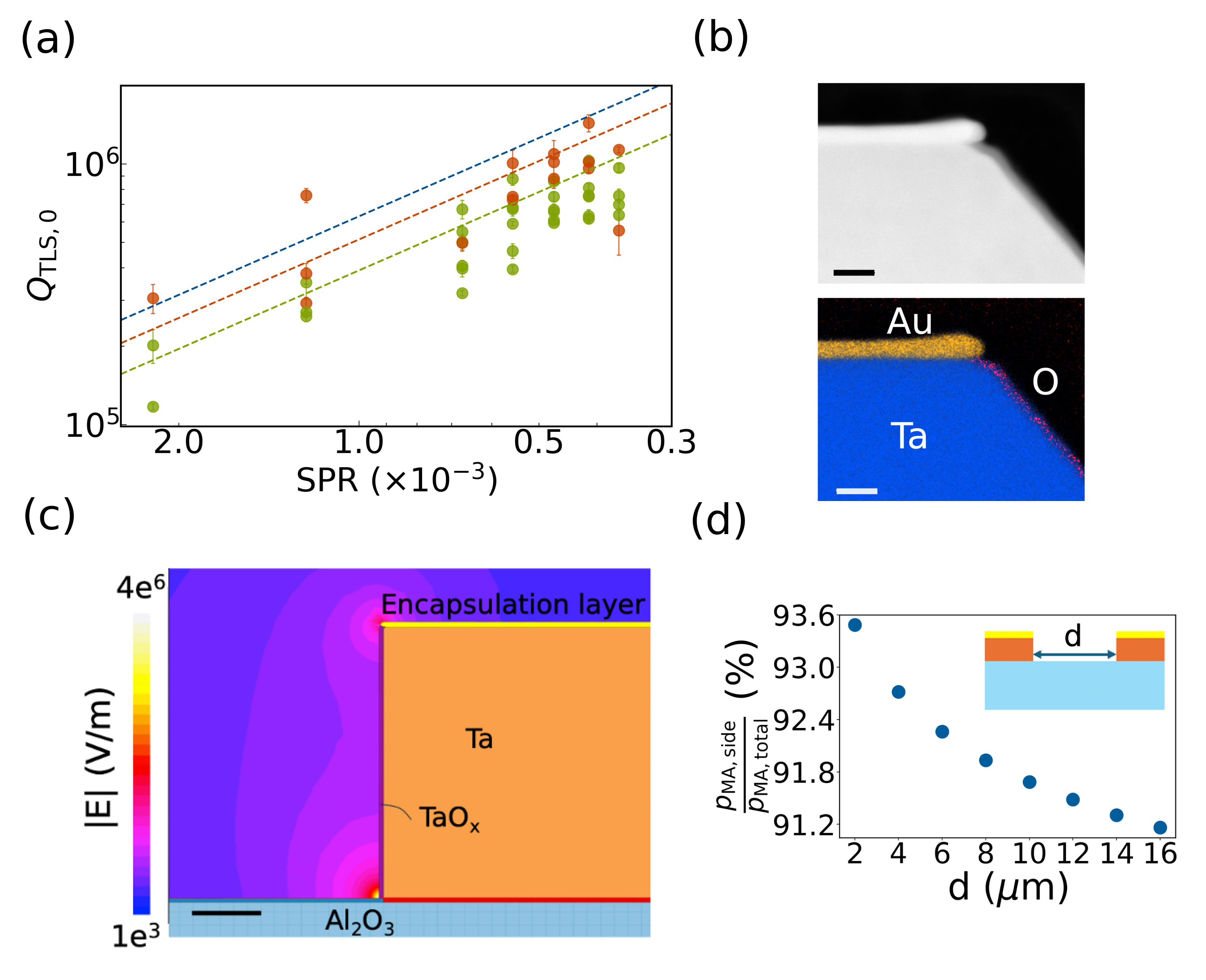}
\caption{\label{fig:fig3} Dielectric loss associated with TLSs of encapsulated Ta resonators. (a) Dependence of $Q_{\text{TLS,0}}$ on SPR. The SPR values are obtained through electromagnetic simulations of the electric field participation at the metal-substrate interface. Dashed lines are fitted surface loss tangents of devices fabricated with bare Ta (blue) from reference \cite{crowley2023}, and those encapsulated with Au (orange) or AuPd (green). Encapsulated devices have a higher fitted surface loss tangent. (b) Cross-sectional STEM-ADF image of an Au-encapsulated Ta resonator (top) and corresponding elemental map extracted from STEM-EDX measurements (bottom). The elemental map is formed by integrating over the Au M (yellow), oxygen K (red), and Ta M (blue) ionization edges. The presence of oxygen on the sidewall indicates tantalum oxide formation. The scale bar in both images is 20 nm long. (c) Dc finite element simulation of the electric field magnitude of a cross section in an encapsulated device. The film is modeled as a 200 nm thick Ta film, with a 3 nm thick sidewall oxide with dielectric constant 10, and a 4 nm thick encapsulating metal layer. Hot spots of electric field intensity are seen at the top and bottom corners of the device. The scale bar is 50 nm long. (d) Ratio of simulated electric field participation of the metal-air interface of sidewalls, $p_\text{MA,side}$, to electric field participation of the total metal-air interface (including sidewalls and top surface), $p_\text{MA,total}$. The ratio is plotted as a function of the distance, $d$, between the center pin and ground plane conductors of the CPW resonators (inset).}
\end{figure}

To characterize the surface and interface losses of encapsulated resonators, we vary their geometry to change the surface participation ratio (SPR) of the electric field energy \cite{wang_surface_2015,crowley2023}. By measuring the temperature-dependent frequency shift, as well as the dependence of internal quality factor on temperature and intracavity photon number, we can also extract the linear absorption due to TLSs, denoted as $Q_{\text{TLS,0}}$, see \cite{crowley2023,our_SI} for details. This extracted parameter captures the contribution of TLSs to dielectric loss. We fit a linear relationship between extracted $Q_\text{TLS,0}$ values and SPR, taking the fitted slope to be the surface TLS dielectric loss tangent [Fig. \ref{fig:fig3}(a), Table \ref{tab:LTs}] \cite{crowley2023,our_SI}. Fitted loss tangents for each film type are also plotted along with loss tangent values for bare Ta devices obtained in our previous work as comparison \cite{crowley2023}. 

\begin{table}[]
    \centering
    \begin{tabular}{ l  cl }
    \hline \hline
         Film & \multicolumn{2}{c}{$\tan \delta$}  \\
         \hline
         Bare Ta & $(15.9 \pm 0.7)$ & $\times  10^{-4}$ \cite{crowley2023} \\
         Au encapsulation & $(19.5 \pm 2)$ & $\times  10^{-4}$ [This work] \\
         AuPd encapsulation & $(25.8 \pm 0.8)$ & $\times  10^{-4}$ [This work] \\
         \hline \hline
    \end{tabular}
    \caption{Surface loss tangents extracted from $Q_{\textrm{TLS,0}}$ values fit to SPR}
    \label{tab:LTs}
\end{table}

Despite the suppression of surface oxides, we do not observe an improvement in the surface TLS loss, and we observe slightly higher surface loss tangents with encapsulated films compared to bare Ta devices. Because our resonators are patterned by a subtractive etch process, the fabrication-exposed sidewalls are not encapsulated and thus grow surface oxide, which can be observed in TEM images [Fig. \ref{fig:fig3}(b)]. To determine the contribution of the sidewall oxide on device performance, we perform dc finite element electric field simulations (Ansys Maxwell) on the cross section of encapsulated devices to determine the contribution of the sidewall oxide to the electric field energy's surface participation. We assume a 3 nm thick oxide layer with dielectric constant 10 on the sidewall and a 4 nm thick encapsulation layer. These two layers make up the the modeled metal-air interface in our devices. We denote the total electric field energy participation in both of these two layers as $p_\text{MA,total}$, and the electric field energy participation in the sidewall oxide as $p_\text{MA,side}$. The sidewall oxide dominates the total participation ratio, accounting for more than 90\% of electric field energy participation along the metal-air interfaces across all device geometries [Fig. \ref{fig:fig3}(d)]. The large participation ratio of the sidewall oxide arises from the 2D geometry of the device, in which the field resides in the gap between metal structures and is concentrated at the corners [Fig. \ref{fig:fig3}(c)].

Beyond the sidewall oxide, the slightly higher measured surface loss tangents of encapsulated devices may arise from increased sidewall roughness. The multilayer structure requires multiple etching steps in the resonator fabrication process, and the additional etching step causes rougher sidewall profiles compared to bare Ta devices \cite{our_SI}. Rougher sidewalls have higher surface area and can concentrate the electric field, and thus increase the surface participation ratio beyond the simulated value, leading to a higher fitted surface loss tangent.

Here we have demonstrated complete suppression of the native surface oxide of Ta using noble metal encapsulation, and we find that these films remain superconducting because of proximitization of the encapsulation layer by the underlying Ta. These encapsulated resonators exhibit performance that is similar to state-of-the-art Ta circuits at mK temperatures. Our ongoing work includes developing fabrication strategies that can achieve total Ta encapsulation to avoid the formation of oxides on the device sidewalls, for example by additive fabrication \cite{tsioutsios2020}. Beyond improving the performance of superconducting qubits, our strategy for superconducting gap engineering via noble metal encapsulation may have applications in on-chip filters for quasiparticle trapping \cite{mcewen2024} and multi-channel kinetic inductance detectors \cite{Vissers_2013, Hu2020}. 

\begin{acknowledgments}
\textit{Acknowledgements.---} We thank Basil Smitham, Kevin Crowley, and Lev Krayzman for helpful discussions. This work was primarily supported by the U.S. Department of Energy, Office of Science, National Quantum Information Science Research Centers, Co-design Center for Quantum Advantage (C2QA) under Contract No. DE-SC0012704. Materials growth and analysis was partially supported by the National Science Foundation (RAISE DMR-1839199). The authors acknowledge the use of Princeton’s Imaging and Analysis Center (IAC), which is partially supported by the Princeton Center for Complex Materials (PCCM), a National Science Foundation Materials Research Science and Engineering Center (MRSEC; DMR-2011750), as well as the Princeton Micro/Nano Fabrication Laboratory. We also acknowledge MIT Lincoln Labs for supplying a traveling wave parametric amplifier.
\end{acknowledgments}

Princeton University Professor Andrew Houck is also a consultant for Quantum Circuits Incorporated (QCI).  Due to his income from QCI, Princeton University has a management plan in place to mitigate a potential conflict of interest that could affect the design, conduct and reporting of this research.

\bibliography{truncated_bib}

\end{document}


\preprint{APS/123-QED}

\title{Supplementary information for ``Eliminating Surface Oxides of Superconducting Circuits with Noble Metal Encapsulation"}

\author{Ray D. Chang}
 \altaffiliation{These authors contributed equally to this work.}
\affiliation{Department of Electrical and Computer Engineering,
Princeton University, Princeton, New Jersey 08540, USA}

\author{Nana Shumiya}
 \altaffiliation{These authors contributed equally to this work.}
\affiliation{Department of Electrical and Computer Engineering,
Princeton University, Princeton, New Jersey 08540, USA}

\author{Russell A. McLellan}
\altaffiliation{Present address: Archer Aviation, San Jose, California 95134, USA}
\affiliation{Department of Electrical and Computer Engineering,
Princeton University, Princeton, New Jersey 08540, USA}

\author{Yifan Zhang}
\affiliation{Department of Electrical and Computer Engineering,
Princeton University, Princeton, New Jersey 08540, USA}

\author{Matthew P. Bland}
\affiliation{Department of Electrical and Computer Engineering,
Princeton University, Princeton, New Jersey 08540, USA}

\author{Faranak Bahrami}
\affiliation{Department of Electrical and Computer Engineering,
Princeton University, Princeton, New Jersey 08540, USA}

\author{Junsik Mun}
\affiliation{Center for Functional Nanomaterials, Brookhaven National Laboratory, Upton, New York 11973, USA}
\affiliation{Condensed Matter Physics and Materials Science Department, Brookhaven National Laboratory, Upton, New York 11973, USA}

\author{Chenyu Zhou}
\affiliation{Center for Functional Nanomaterials, Brookhaven National Laboratory, Upton, New York 11973, USA}

\author{Kim Kisslinger}
\affiliation{Center for Functional Nanomaterials, Brookhaven National Laboratory, Upton, New York 11973, USA}

\author{Guangming Cheng}
\affiliation{Princeton Materials Institute, Princeton University, Princeton, New Jersey 08540, USA}

\author{Alexander C. Pakpour-Tabrizi}
\affiliation{Department of Electrical and Computer Engineering,
Princeton University, Princeton, New Jersey 08540, USA}

\author{Nan Yao}
\affiliation{Princeton Materials Institute, Princeton University, Princeton, New Jersey 08540, USA}

\author{Yimei Zhu}
\affiliation{Condensed Matter Physics and Materials Science Department, Brookhaven National Laboratory, Upton, New York 11973, USA}

\author{Mingzhao Liu}
\affiliation{Center for Functional Nanomaterials, Brookhaven National Laboratory, Upton, New York 11973, USA}

\author{Robert J. Cava}
\affiliation{Department of Chemistry,
Princeton University, Princeton, New Jersey 08540, USA}

\author{Sarang Gopalakrishnan}
\affiliation{Department of Electrical and Computer Engineering,
Princeton University, Princeton, New Jersey 08540, USA}

\author{Andrew A. Houck}
\affiliation{Department of Electrical and Computer Engineering,
Princeton University, Princeton, New Jersey 08540, USA}

\author{Nathalie P. de Leon}
 \altaffiliation{npdeleon@princeton.edu}
\affiliation{Department of Electrical and Computer Engineering,
Princeton University, Princeton, New Jersey 08540, USA}

\date{\today}
\maketitle
\tableofcontents

\renewcommand{\thefigure}{S\arabic{figure}}

\section{\label{sec:fab}Sample fabrication} 
\subsection{Film growth}

For the films used for surface oxide characterization with XPS, we deposited Nb-seeded $\alpha$-Ta in the $\langle110\rangle$ orientation [Fig. \ref{fig:sup_xrd}] using a dc magnetron sputtering process \cite{Face_Prober_1987} at room temperature in an AJA ATC Orion sputtering system with a base pressure of 5$\times 10^{-9}$ Torr. For all sputtering processes, Ar is flowed at a rate of 26 sccm and held to a chamber pressure of 3 mTorr. Both the Nb and the Ta depositions are sputtered at 250W on a dc power supply, with respective deposition rates of 1.5 and 2 $\textup{~\AA}$/s. The Ta deposition is followed by an in situ deposition of Au or AuPd. The AuPd sputter target is in a 48\% to 52\% weight ratio of Au to Pd. Au and AuPd are sputtered at 10W on an rf power supply, with deposition rates of around .075 $\textup{~\AA}$/s.  To calibrate the precise thicknesses of the encapsulation layers, the deposition rates of Au and AuPd are first found by measuring an etched step with a Tencor P-17 profilometer on multiple encapsulation layer depositions in which the deposition time was varied and the deposition power is held constant. The deposition rate is assumed to be constant in time, and is then extracted by fitting the thicknesses to a constant deposition rate. The deposition time is then adjusted to reach nominal targeted deposition thicknesses.

For fabricated resonators, we used films deposited by Star Cryoelectronics. The films were deposited on c-plane HEMEX grade sapphire wafers, which are first cleaned by a 20 minute piranha acid (2:1 ratio of sulfuric acid and hydrogen peroxide) treatment and an oxygen plasma treatment before deposition. 200 nm of $\alpha$-Ta is deposited on the substrate at 500$\degree$C with a dc magnetron sputtering process. The encapsulating Au and AuPd is then sputtered \text{in situ} following the Ta deposition without breaking vacuum. The encapsulation layer is around 4-5 nm in thickness. 

To fabricate resonators with Au encapsulation layers with thicknesses up to 323 nm, we sputtered additional Au on top of the Star Cryoelectronics Au encapsulated films (after a piranha solution clean of the surface) in an AJA ATC Orion sputtering system, using an rf power supply at 250W with a deposition rate of around 2 $\r{A}$/s.

\subsection{Device fabrication}

After deposition, the encapsulated films are primed with hexamethyldisilane (HMDS) in a YES vapor prime oven at 148$\degree$C to promote resist adhesion to the noble metal surface. Then, AZ1518 is spun on the film at 4000 rpm for 45 s and soft baked at 95$\degree$C for 1 minute. The photoresist is then patterned using a Heidelberg DL66+ laser writer, using a 10 mm write head, with a 50\% attenuator, intensity setting of 30\%, and a focus offset setting of 10\%. Following patterning, the photoresist is then baked at 110$\degree$C for 2 minutes and then cooled on a metal plate for 1 minute. The photoresist is then developed in AZ300MIF solution for 90 s and rinsed in deionized (DI) water for 30 s. The patterned photoresist is then used as a mask for subsequent etch steps.

\begin{figure*}[b]
 \includegraphics[width=\linewidth]{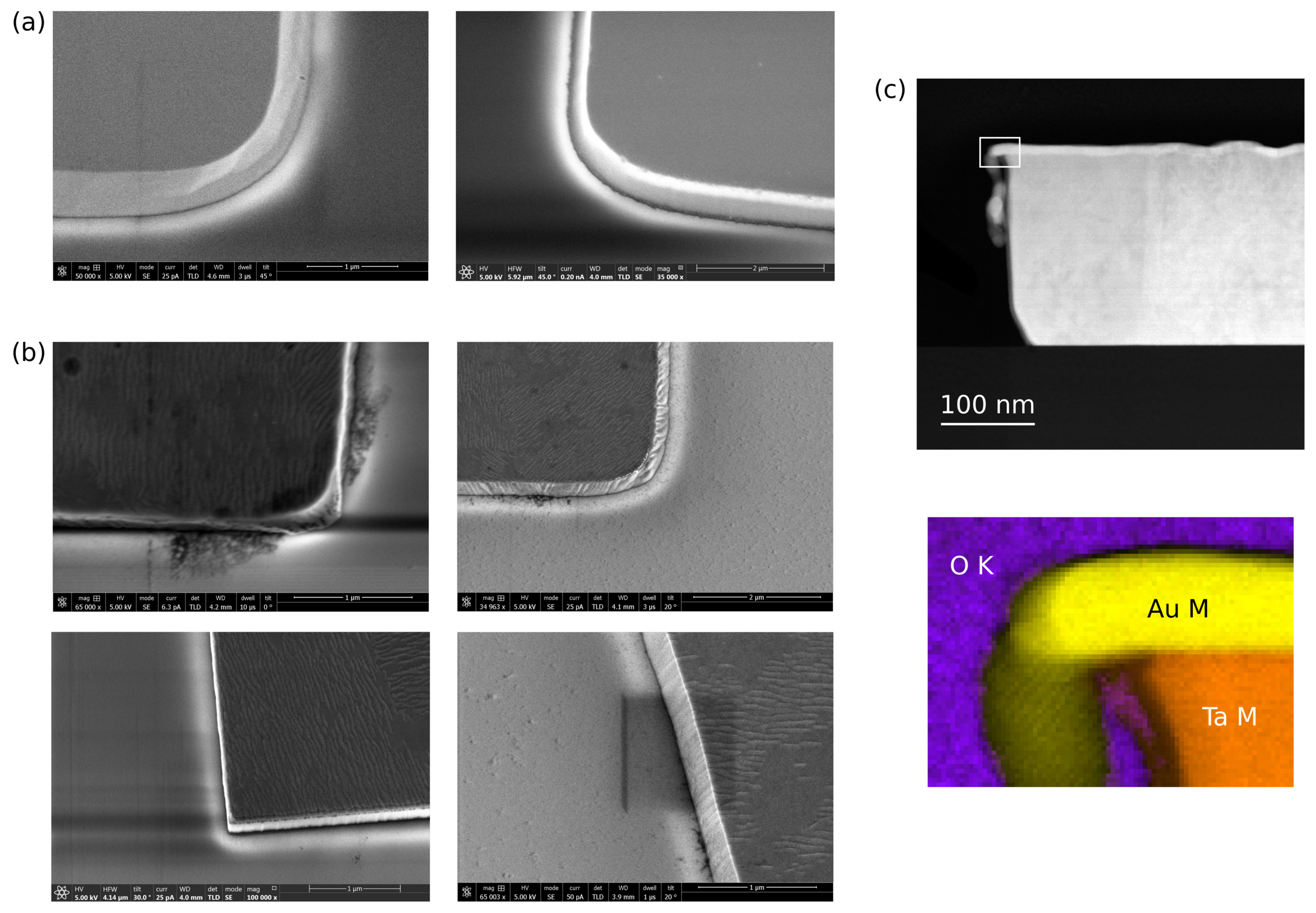}
\caption{\label{fig:sup_fab}  (a) Scanning electron microscope images of bare Ta devices patterned with chlorine-based dry chemical etch (left) and Ta wet etchant (right). The surface TLS dielectric loss tangent of these devices were measured in our previous work \cite{crowley2023}. (b) Scanning electron microscope images of various Au/AuPd encapsulated devices patterned with aqua regia and wet etchant (top left), aqua regia and fluorine-based dry chemical etch (bottom left) and aqua regia and chlorine-based dry chemical etch (top right and bottom right). (c) High angle annular dark field (HAADF) scanning TEM image of the cross-section of top corner of an encapsulation device (top) and STEM-EELS maps of gold, tantalum, and oxygen corresponding to the white boxed area in the TEM image.}
\end{figure*}

Next, the Au or AuPd layer is etched in a dilute aqua regia solution. Dilute aqua regia is mixed in a 2:3:1 ratio of DI water:HCl:HNO$_3$, and the Au and AuPd films are etched at approximately 8 $\textup{~\AA}$/s. We note that the encapsulated devices show more sidewall roughness than bare Ta films when undergoing identical fabrication steps except for the added aqua regia etch on encapsulated devices [Fig. \ref{fig:sup_fab}]. 

The underlying tantalum is then etched with one of three etch types (these etch steps were used in our previous work \cite{crowley2023}). One type is a wet chemical etch, 1:1:1 ratio of HF:HNO$_3$:H$_2$O (Transene Tantalum Etchant 111), in which a sample is swirled for 21 s before being rinsed in 3 cups of DI water and 1 cup of 2-propanol and then blow dried in N$_2$. The second etch type is a chlorine-based dry chemical etch in an inductively coupled plasma reactive ion etcher (PlasmaTherm Takachi). The etching parameters for the chlorine dry etch are as follows: ambient pressure of 5.4 mTorr, chlorine flow rate of 5 sccm, argon flow rate of 5 sccm, rf power of 500 W, and bias power of 50 W, which results in an etch rate of approximately 100 nm/min. The third etch type is a fluorine-based dry etch, using the same reactive ion etcher as the chlorine etch, with parameters: ambient pressure 50 mTorr, CHF$_3$ flow rate 40 sccm, SF$_6$ flow rate 15 sccm, Ar$_3$ flow rate 10 sccm, rf power of 100 W, and bias power of 100 W.

After both etch steps are complete, the photoresist mask is stripped in a Remover PG bath at 80$\degree$C for at least 1 hour, followed by a 2 minute each sonication in toluene, acetone, and 2-propanol. The chips are then further cleaned with a 20 minute piranha solution dip. Finally, the devices are wirebonded and packaged in a commercial microwave package (QDevil QCage.24).

\section{\label{sec:measurement} Measurement apparatus} 

All devices were measured in a BlueFors XLD dilution refrigerator with a base mixing chamber temperature of approximately 10 mK. Our measurement setup is described in detail in \cite{crowley2023}. Some of the measurements were also conducted with a Xilinx RFSoC using the Quantum Instrumentation Control Kit (QICK) firmware \cite{rfsoc}.

\section{\label{sec:xrd}Crystallinity of tantalum films} 

\begin{figure*}[h]
\includegraphics[width=\linewidth]{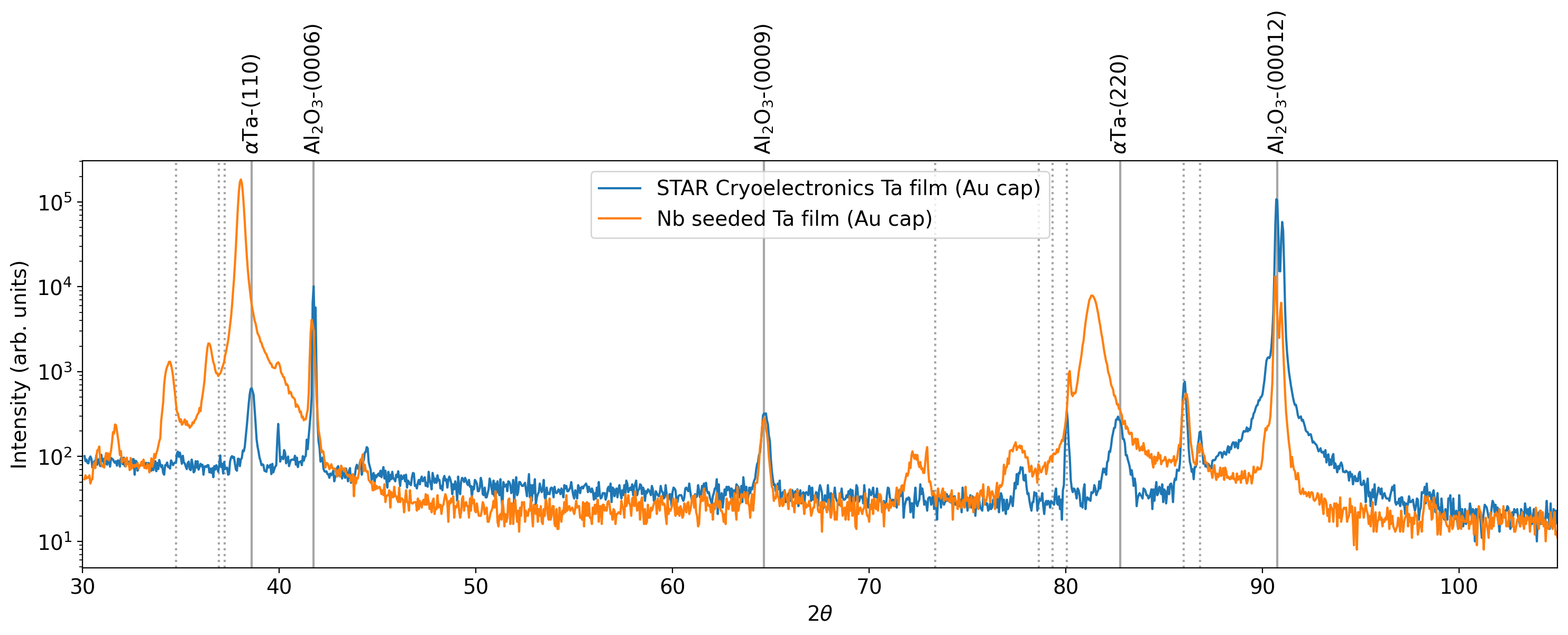}
\caption{\label{fig:sup_xrd} XRD spectrum of encapsulated tantalum films. XRD spectrum of a bare sapphire wafer is also included for comparison.}
\end{figure*}

We use X-ray diffraction (XRD) on a Bruker D8 Discover Diffractometer with a K-$\alpha$ X-ray source to study the crystal structure of Ta in our encapsulated films. For Au and AuPd encapsulated films used for resonator fabrication, the XRD spectra contain clear peaks corresponding to $\alpha$-Ta $\langle110\rangle$ and $\langle220\rangle$ [Fig. \ref{fig:sup_xrd}] \cite{Gladczuk2004}. We also detected several peaks corresponding to sapphire \cite{nakagomi2012,mukherjee2012}. We do not detect any $\beta$-Ta peaks, which provide evidence that the Ta is uniformly in the $\alpha$ phase. The unlabeled dotted lines are secondary echo XRD peaks corresponding to the copper K-$\beta$, tungsten L-$\alpha$1, and tungsten L-$\alpha$2 wavelengths. These wavelengths are emitted from the X-ray source with less intensity, and hence cause visible peaks offset from primary peaks. Other few unassigned small peaks are considered to originate from contamination, instrumental artifacts or impurities and defects in the films. For films used for surface oxide characterization with XPS, we also observe $\alpha$-Ta $\langle110\rangle$ peaks that are slightly shifted to smaller angles [Fig. \ref{fig:sup_xrd}], which we hypothesize is due to different strain conditions resulting from the room temperature growth. Atomic force microscopy (AFM) scans of these films' surfaces are consistent with the surface morphology of $\langle110\rangle$ oriented films [Fig. \ref{fig:sup_afm_au_aupd}], and DC transport measurements show a superconducting $T_c$ matching that of the $\alpha$ phase [Fig. \ref{fig:sup_dctransport_nb-ta}].

\section{\label{sec:afm}Surface morphology of encapsulated films} 

\begin{figure*}[h]
 \includegraphics[width=\linewidth]{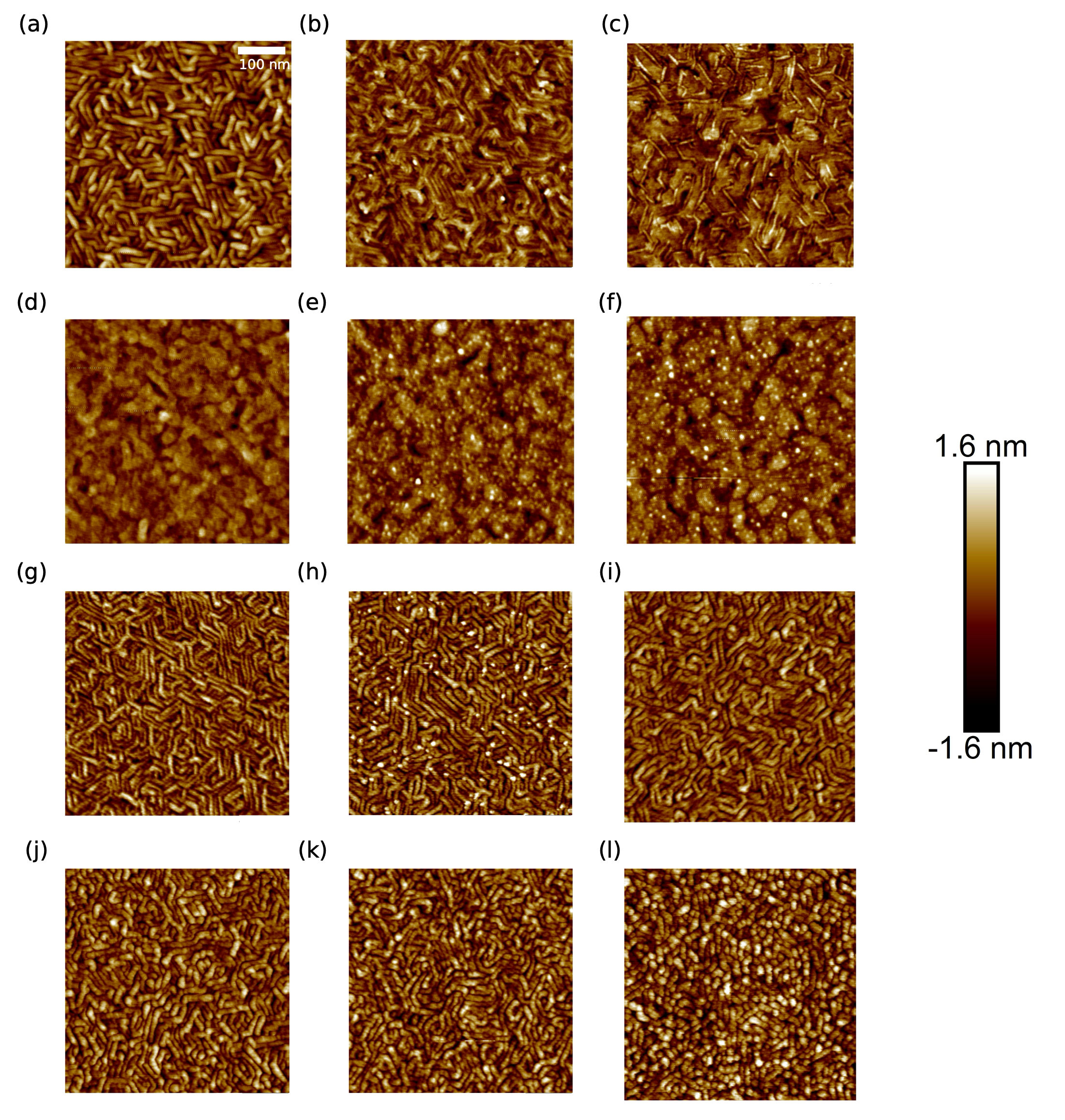}
\caption{\label{fig:sup_afm_au_aupd} AFM scans of (a): bare Ta surface, and Ta surfaces covered with (b-f): 1, 2, 3, 4, 5 nm of Au, respectively, and (g-l): 0.5, 1, 1.5, 2, 3, 4 nm of AuPd, respectively. The scale bars are the same for each figure.
}
\end{figure*}

We studied the surface morphology of Ta films encapsulated with Au or AuPd using AFM, taken on a Bruker Dimension ICON3 AFM. On bare Ta, the facets of the $\langle110\rangle$ surface are clearly visible, whereas they become less visible in scans on films with progressively increasing amounts of Au or AuPd [Fig. \ref{fig:sup_afm_au_aupd}]. Additionally, our extracted surface roughness values on fully encapsulated films in which the oxide growth is suppressed is much less than the encapsulation layer thickness, indicating that we have a homogeneous encapsulation of the Ta surface [Table \ref{tab:afm_au_roughness}].

\begin{table}[]
    \centering
    \begin{tabular}{ c || c | c || c | c}
        \hline \hline 
        Encapsulation& \multicolumn{2}{c||}{Au} & \multicolumn{2}{c}{AuPd} \\
         thickness (nm)& $R_q$ (nm) & $R_a$ (nm) & $R_q$ (nm) & $R_a$ (nm)\\
         \hline
         0  & $0.47$ & $0.37$ & $0.47$ & $0.37$    \\
         0.5 & - & - & $0.25$ & $0.20$ \\         
         1 & $0.43$ & $0.34$ & $0.52$ & $0.41$\\
         1.5 & - & - & $0.50$ & $0.40$ \\
         2 & $0.43$ & $0.34$ & $0.48$ & $0.38$\\
         3 & $0.30$ & $0.24$ & $0.51$ & $0.40$\\
         4 & $0.38$ & $0.30$ & $0.58$ & $0.46$ \\
         5 & $0.40$ & $0.30$ & - & -\\
         \hline \hline

    \end{tabular}
    \caption{Surface average roughness ($R_q$) values and root mean square roughness ($R_a$) values extracted from AFM scans on $\alpha$-$\langle110\rangle$ Ta films encapsulated with varying amounts of Au or AuPd.}
    \label{tab:afm_au_roughness}
\end{table}

\section{\label{sec:xps}XPS characterization} 
\subsection{Experimental and analysis details}
\begin{figure}[ht]
 \includegraphics[width=0.5\linewidth]{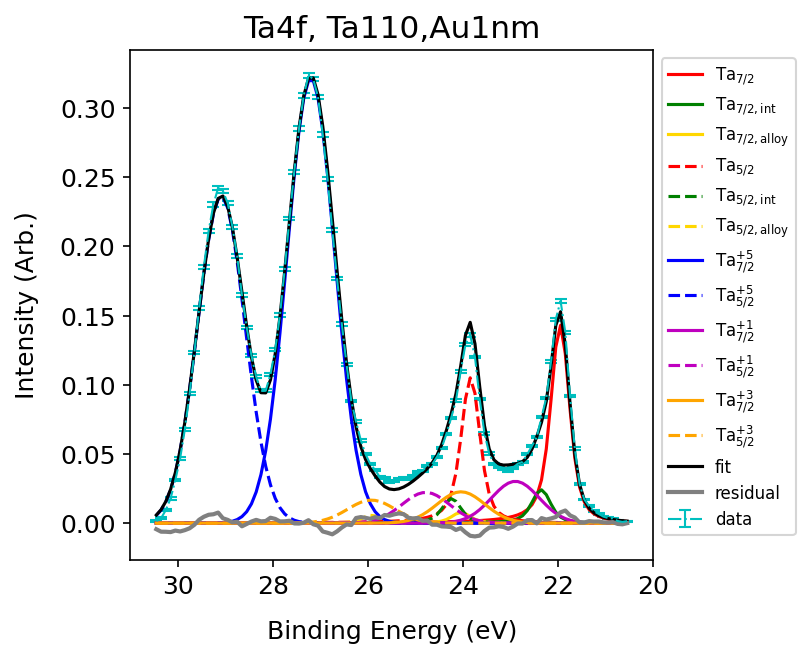}
\caption{\label{fig:sup_xps_1} Shirley background subtracted Ta4$f$ binding energy XPS spectrum fit to a multi-component model, following \cite{mclellan_xps2023}. The film is $\alpha$-Ta encapsulated with nominally 1 nm of Au.}
\end{figure}

\begin{figure}[ht]
 \includegraphics[width=0.4\linewidth]{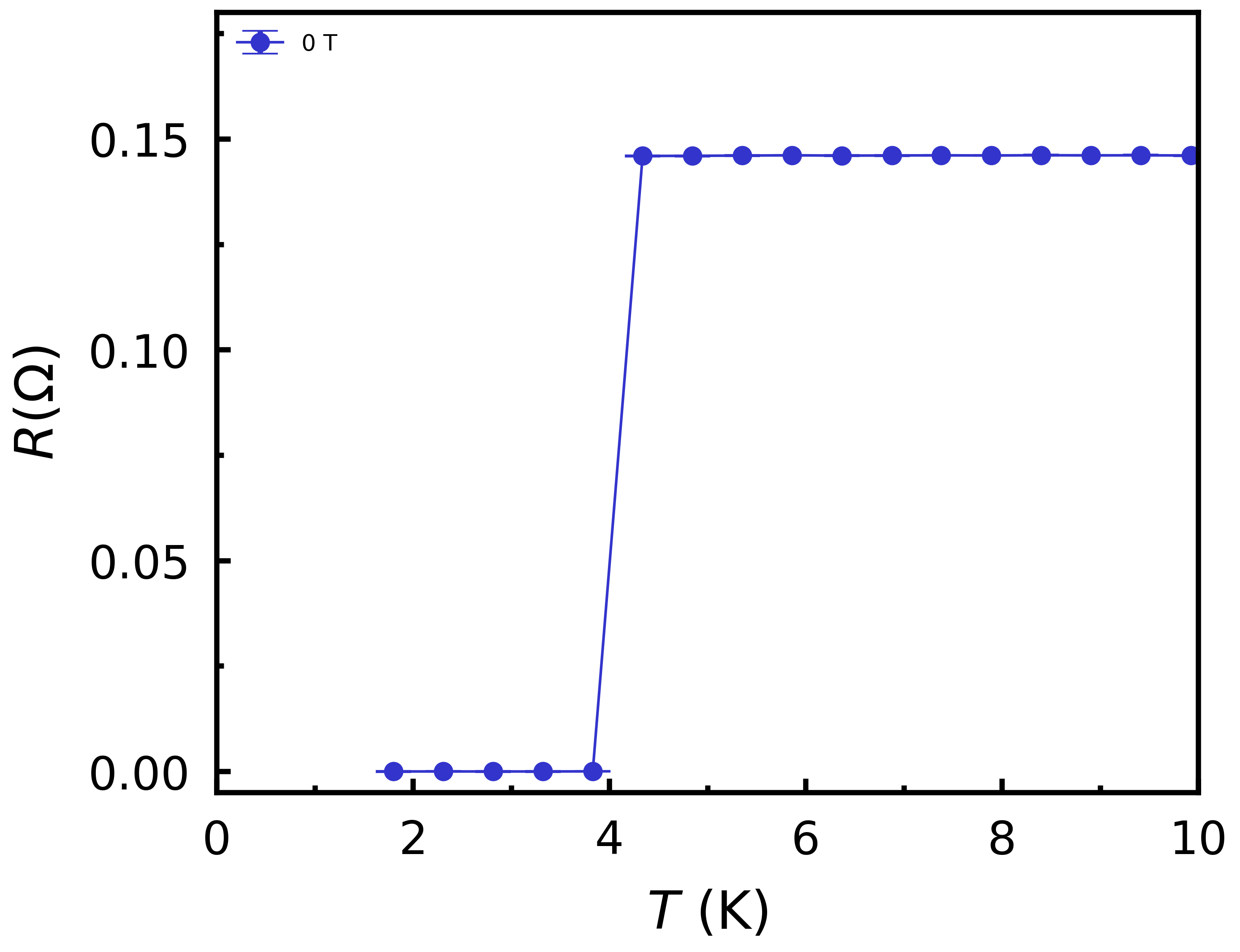}
\caption{\label{fig:sup_dctransport_nb-ta} DC transport measurement on a Nb-seeded $\langle110\rangle$ $\alpha$-Ta film.}
\end{figure}

All XPS measurements are taken on a ThermoFisher K-Alpha XPS spectrometer with an aluminum K$\alpha$ X-ray source. We take XPS scans of the Ta4f and Au4f binding energy spectra with an energy step size of 0.1 eV, a dwell time of 100 ms, and averaged the spectra over 20 sweeps. To calculate the Ta$_2$O$_5$ fraction, a Shirley background is first subtracted from the Ta4f peaks, which are then fit to a multi-component model for all of the observed peaks \cite{mclellan_xps2023} [Fig. \ref{fig:sup_xps_1}]. The tantalum pentoxide fraction, $p$, is extracted as a ratio of metal and Ta$^{5+}$ peak intensities, weighted by the density of Ta atoms within the respective species. Explicitly, $p=\frac{I_o/N_o}{I_o/N_o+I_m/N_m}$, where
$I_o$ and $I_m$ denote the intensity of the Ta$^{5+}$ and metallic Ta layers, and $N_o$ and $N_m$ denote the Ta atom density of the Ta$^{5+}$ and metallic Ta layers. We focus on the Ta$^{5+}$ peak intensities as it is the dominant oxide species.

\subsection{Encapsulation layer dependent oxide thickness}
We observe an increase in the tantalum oxide intensity when we deposit less than 2.5 (1) nm, nominally, of Au (AuPd) on the Ta surface [Fig. 1(d)]. In this regime, the encapsulation material does not form a complete, continuous layer [Appendix \ref{sec:afm}]. We hypothesize that the presence of small Au islands catalyzes additional oxide growth by promoting both oxygen adsorption to the tantalum surface and causing oxidation-favorable charge transfer \cite{hu2021, yang2022}. This leads to a larger oxide intensity. Once the surface is completely encapsulated we observe that the oxide intensity drops to zero. 

\section{\label{sec:transport}DC resistivity measurements}

We used a Quantum Design PPMS Dyna Cool instrument to measure the dc resistance as a function of temperature for bare Ta and Nb-seeded Ta films. We used the four-point probe resistance method with a current of 100 $\mu$A. The rate for our measurements was 2 K per minute; we stabilized and recorded the resistance value either at every 0.1 K or every 0.5 K depending on the measurement. We extract $T_c$ at 50\% of the superconducting transition step. For bare Ta, we found a $T_c$ of $4.4\pm.1$ K. For Nb-seeded Ta, we found a $T_c$ of $4.083\pm.25$ K, which is close to the expected value for an $\alpha$-Ta film.

\section{\label{sec:tls}Characterizing the surface TLS contribution to dielectric loss and the superconducting energy gap}
We follow the methodology of \cite{crowley2023} to extract the contribution of surface TLSs to dielectric loss and the superconducting energy gap in our devices. The different sources of loss in our devices can be disentangled via a temperature-dependent frequency shift measurement or by characterizing internal quality factors across a wide range of intracavity photon number and temperature. By fitting the frequency shift or quality factor data to a combined TLS-quasiparticle model, the linear absorption due to TLSs can be characterized in a fit parameter we denote as $Q_\text{TLS,0}$. Explicitly, we model the TLS losses as \cite{crowley2023}
\begin{equation}
Q_{\text{TLS}}(\bar{n},T) = Q_{\text{TLS,0}}\frac{\sqrt{1+(\frac{\bar{n}^{\beta_2}}{DT^{\beta_1}})\tanh{(\frac{\hbar\omega}{2 k_b T})}}}{\tanh{(\frac{\hbar\omega}{2 k_b T})}}
\end{equation}
where $\omega$ is the center angular frequency of the resonator; $T$ is the temperature; $\bar{n}$ is the intracavity photon number; $\hbar$ is the reduced Plank's constant; $k_B$ is the Boltzmann constant; and $D, \beta_1, \beta_2$ are additional fit parameters characterizing TLS saturation. Equivalently, the frequency shift due to TLSs is modeled as \cite{crowley2023,gao_jiansong_physics_2008}
\begin{equation}
\left(\frac{\delta f(T)}{f}\right)_\text{TLS} = \frac{1}{\pi Q_\text{TLS,0}}\text{Re}\left[\Psi\left(\frac{1}{2}+i\frac{\hbar\omega}{2\pi k_B T}\right)-\ln\left(\frac{\hbar\omega}{2\pi k_B T}\right)\right]
\end{equation}
where $\Psi$ is the complex digamma function.

To isolate the contribution of surface TLSs to dielectric loss, we tune the SPR of our devices by tuning the gap spacing between the center pin conductor and the ground plane in our CPW resonators. The extracted $Q_{\text{TLS,0}}$ scales with SPR [Fig. 3(b)] and can be modeled as arising from two components: a surface TLS bath that scales with SPR and a bulk TLS bath that is SPR independent. The surface TLS dielectric loss tangents for each kind of surface preparation (bare Tantalum, Au-encapsulated, and AuPd-encapsulated) can be extracted by fitting these two components to the extracted $Q_{\text{TLS,0}}$ values across all devices [Table I]. Following convention, we choose the metal-substrate interface participation as the independent variable to fit our quality factors to in order to extract dielectric loss tangents \cite{wang_surface_2015,crowley2023}. For our CPW resonators, the surface TLS component dominates the extracted $Q_{\text{TLS,0}}$ values as the SPR independent bulk loss tangent is much lower \cite{crowley2023}.

The effective superconducting energy gap of the films that our resonators are fabricated from can also be extracted via an internal quality factor measurement or by a frequency shift measurement. The effects of thermal quasiparticles on the internal quality factor of our resonators is modeled as \cite{crowley2023,gao_jiansong_physics_2008}

\begin{equation}
Q_\textrm{QP}(T) = A_{\textrm{QP}} \frac{e^{\Delta /k_B T}}{\sinh \left( \frac{\hbar \omega}{2 k_B T} \right) K_0 \left( \frac{\hbar \omega}{2 k_B T} \right)}
\end{equation}
where $A_{\textrm{QP}}$ is an overall amplitude proportional to the kinetic inductance ratio; $\Delta$ is the effective superconducting energy gap; $K_0$ is the zeroth order modified Bessel function of the second kind; $k_B$ is the Boltzmann constant. Equivalently, the frequency shift induced by thermal quasiparticles is modeled as \cite{crowley2023,gao_jiansong_physics_2008}
\begin{equation}
\left(\frac{\delta f(T)}{f}\right)_\text{QP} = -\frac{\alpha}{2}\left(\frac{|\sigma(0,\omega)|}{|\sigma(T,\omega)|}\sin(\phi(T,\omega)-1\right)
\end{equation}
where $\alpha$ is the kinetic inductance fraction; $\sigma$ is the superconductor complex conductivity; and $\phi$ is the phase of the superconductor complex conductivity.

Our quality factor and frequency shift measurements are simultaneously fit to both the TLS and quasiparticle components, allowing for the extraction of both the TLS contribution to dielectric loss and the effective superconducting energy gap.

\section{\label{sec:usadel}Modeling superconducting gap of encapsulated films with Usadel equations}
\subsection{Overview}

The Usadel equation is the standard formalism for describing superconductor-normal metal heterostructures \cite{belzig98, martinis_2000}. It describes the complex, position- and energy-dependent pairing angle \(\theta(x,E)\), where \(x\) is the position and \(E\) is the energy.
\(\theta(x,E)\) parameterizes the two-by-two retarded Green's function of the electron. \(\theta=0\) in the normal state, while a non-trivial \(\theta\) indicates that superconducting order is present. In the bulk of the superconductor, \(\theta\) takes the BCS value of \(\theta=i \, \text{arctan}(\Delta_0/E)\) where \(\Delta_0\) is the bulk order parameter.

The Usadel equation is a nonlinear diffusion equation of \(\theta(x,E)\) in the normal metal with thickness \(d_N\) and in the superconductor with thickness \(d_S\) (See Fig. 2(b) for the schematics):
\begin{align}
\frac{\hbar D}{2} \frac{\partial^2 \theta}{\partial x^2} + iE\sin(\theta)+\Delta(x)\cos(\theta)=0
\end{align}
where \(D\) is the diffusion coefficient and \(\Delta(x)\) is the local order parameter.
\(\Delta(x)=0\) in the normal metal, while \(\Delta(x)\) is related to \(\theta(x,E)\) in the superconductor via the following self-consistent equation.
\begin{align}
    \Delta(x) = N_S V \int_0^{\hbar \omega_D} \tanh (\frac{E}{2 k_B T}) \text{Im} [\sin(\theta(x))] dE
\end{align}
Where \(N_S\) is the normal-state DOS at the Fermi surface of the superconductor, \(V\) is the pairing strength, \(\hbar \omega_D\) is the Debye energy, and \(T\) is the temperature.
The Usadel equation admits the following boundary condition at the metal-superconductor interface. As our convention, we set the metal-superconductor interface as \(x=0\).
\begin{align}\label{eq:real_usadel}
    \sigma_{N,S} \left(\frac{\partial \theta_{N,S}}{\partial x}\right)_{x=0} = \frac{G_{\text{int}}}{A} \sin (\theta_S(0,E)-\theta_N(0,E))
\end{align}
Where \(\sigma_{N,S}\) and \(\theta_{N,S}\) denote the conductivity and the pairing angle in the normal metal and the superconductor. \(G_{\text{int}}/A\) denotes the interface conductance per unit area.
Similarly, the metal-air and the superconductor-air boundaries have the following boundary condition where \(G_{\text{int}}=0\).
\begin{align}\label{eq:real_usadel_bdry}
    \left(\frac{\partial \theta_{N,S}}{\partial x}\right)_{x=-d_N, d_S} = 0
\end{align}
Finally, the local spectral density \(N(x,E)\) can be computed from \(\theta(x,E)\) once it is known.
\begin{align}\label{eq:real_usadel_dos}
    N(x,E) = N_{N,S} \text{Re} [\cos (\theta)]
\end{align}
where \(N_{N,S}\) is the normal-state DOS at the Fermi surface of the normal metal or the superconductor. 
Because of the proximity effect, $N(x,E)$ becomes suppressed below an energy called the minigap (See Fig. 2(b)). In general, the minigap is a function of the position and takes the smallest value at the metal-air interface. We take this smallest gap and fit to the experimental value.

The above set of equations can be solved self-consistently to find the minigap. In practice, the Usadel equation in the current form is highly numerically unstable.
Therefore, following \cite{belzig96,belzig98}, we solve for \(\Delta(x)\) self-consistently by solving an equivalent imaginary-time problem. Then we solve Eqns. \ref{eq:real_usadel}, \ref{eq:real_usadel_bdry}, and \ref{eq:real_usadel_dos} once to obtain the spectral density $N(x,E)$.

In the imaginary-time formalism, we solve a \textit{real} pairing angle \(\tilde{\theta}(x,E)\). 
\(\tilde{\theta}=0\) in the normal metal while \(\tilde{\theta}=\text{arctanh}(\Delta_0 / E)\) in the bulk of the superconductor. 
The Usadel equation under the imaginary-time formalism now reads as
\begin{align}
    \frac{\hbar D}{2} \frac{\partial^2 \tilde{\theta}}{\partial x^2} -E\sin(\tilde{\theta})+\Delta(x)\cos(\tilde{\theta})=0
\end{align}
And the order parameter \(\Delta(x)\) is related to \(\tilde{\theta}\) via
\begin{align}
    \Delta(x) = 2 \pi k_B T N_S V \sum_{\omega_n}^{\omega_D} \sin(\tilde{\theta}(x,\hbar \omega_n))
\end{align}
where \(\omega_n= \pi k_B T (2n+1) / \hbar \) denotes the Matsubara frequencies.
\subsection{Fitting procedure}
We solve the Usadel equation self-consistently and numerically as previously described. 
We then then fit to the experimentally measured minigap at various Au thickness \(d_N\) [Fig. 2]. To measure \(D_{N,S}\), we perform DC transport measurements of conductivity of the Au and the Ta (see Appendix \ref{sec:transport}) and deduce \(D_{N,S}\) as follows.
To measure the conductivity of Au on top of the Ta, we deposit Au at various thicknesses and measure the in-plane electrical conductivity of the resulting heterostructures. As the length of the structure is much greater than the thickness of either layer, we extrapolate the Au conductivity with a parallel resistor model, with one resistor being the Ta layer and the other being the Au layer.

We then relate the conductivity \(\sigma_{N,S}\) to \(D_{N,S}\) under the free electron approximation \cite{ashcroft_mermin}. The mean free path \(\lambda_{N,S}\) is then related to the conductivity via \(\lambda_{N,S}=\sigma_{N,S} m_e v_F  / (n_{N,S} q^2)\), where $q$ is the electron charge, \(m_e\) is the electron mass and \(n_{N,S}\) is the electron density in Au or Ta.
After that, the diffusion coefficient can be computed with \(D_{N,S}=(v_{F})_{N,S}\lambda_{N,S}/3\), where \((v_{F})_{N,S}\) denotes the Fermi velocity of Au or Ta.
We take values of \(n_{N}\) and \((v_{F})_{N}\) from \cite{ashcroft_mermin} and take values of \(n_{S}\) and \((v_{F})_{S}\) from \cite{bobrov2015point}.
We set the Ta thickness to \(d_S=200 \, \text{nm}\) and measure \(D_{S,N}\) and \(\Delta_0\) experimentally. 
After that, the only free parameter in this model is the interface conductivity \(G_{\text{int}}\) which we treat as a fitting variable.

\begin{figure}[htbp]
 \includegraphics[width=0.6\linewidth]{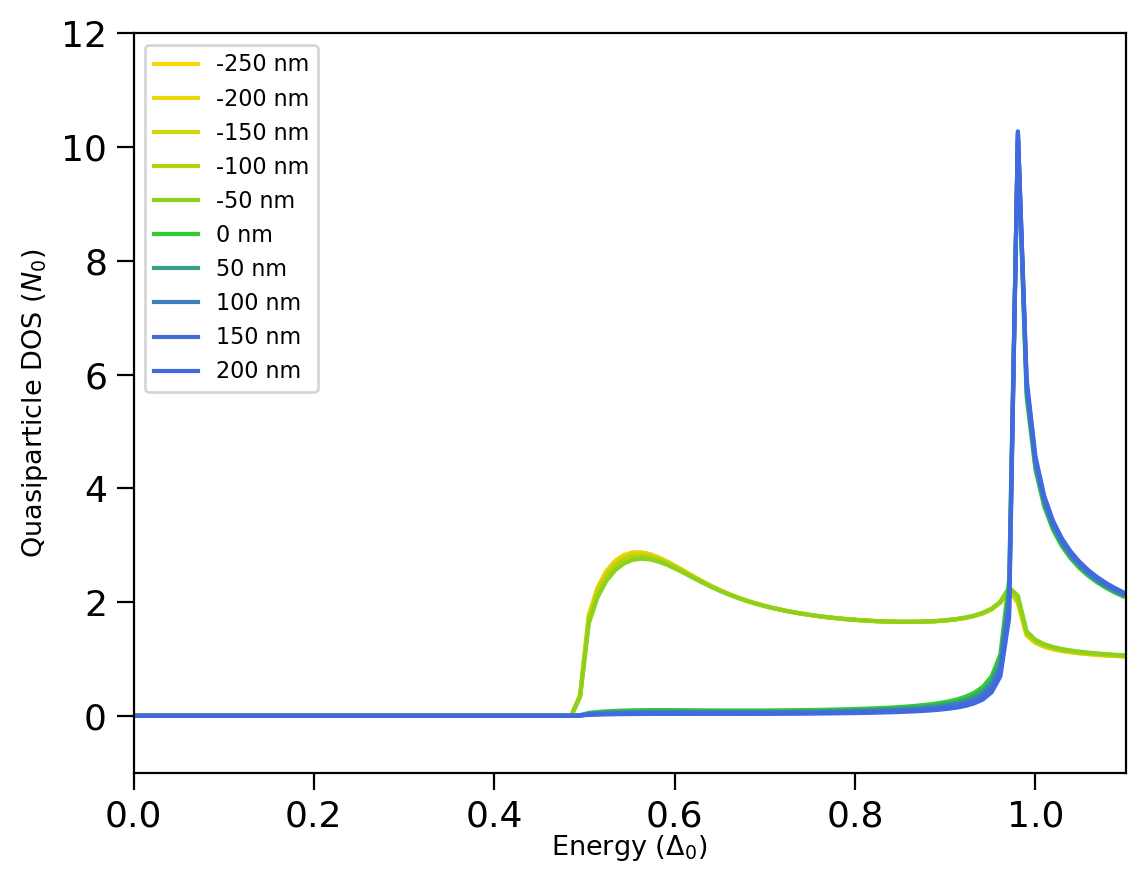}
\caption{\label{fig:collapsed-DOS} Quasiparticle DOS for different depth positions in a Ta-Au bilayer. The interface is at 0 nm, and Au and Ta are at negative and positive depth values, respectively.}
\end{figure}

We note that the free electron approximation is a good approximation of Au because it has one valence electron per unit cell and has a roughly isotropic Fermi surface.
On the other hand, Ta is multi-valent and has convoluted Fermi surfaces. Nevertheless, \(D_{S}\) is not a sensitive parameter in this model, so approximating Ta under the free electron approximation does not significantly modify the result.
Finally, \(G_{\text{int}}\) can be converted to a normalized transparency coefficient, as reported in the main text
\begin{align}
t=\frac{(G_{\text{int}}/A)}{(2 G_q / (\lambda_S/2)^2)}
\end{align}
Where \(G_q=q^2/h\) denotes the conductance quantum and \(\lambda_S\) denotes the Fermi wavelength of the superconductor. We summarize the model parameters in Table \ref{tab:usadel_params}. Notice that Au has a very high scattering length at low temperature, comparable to the Au thickness. This results in the relatively consistent local spectral density everywhere in the Au (Fig. 2(b), Fig. \ref{fig:collapsed-DOS}). The gap is mostly controlled by the low interface transparency which gives rise to the discontinuity in the local spectral density across the interface.

\begin{table}
\caption{\label{tab:usadel_params} Parameters of the Usadel equation}
\begin{ruledtabular}
    \begin{tabular}{llll}
    $G_{\text{int}}/A$ (S/$\mu$m$^2$)     & $271^{+191}_{-90}$         & $t$              & $0.24^{+.17}_{-.08}$ \\
    $D_N$ (cm$^2$/s) & 622 $\pm$ 249         & $\lambda_N$ (nm) & 133 $\pm$ 53.3 \\
    $D_S$ (cm$^2$/s) & 19.2 $\pm$ .6        & $\lambda_S$ (nm) & 5.57 $\pm$ 0.18\\
    $T_c$ (K)        & 4.4 $\pm$ .1        &                  &     
    \end{tabular}
\end{ruledtabular}
\end{table}

Note that the error bars in Fig. 2(c) are derived from the statistical error of fitting the superconducting gap to the resonator measurements, and they significantly underestimate the true statistical fluctuations originating from other sources such as sample variations. This is evident from the first few data points at low thickness of Au where they behave non-monotonically and spreads out by roughly 0.5 K. Therefore, it is hard to quantify the confidence of the fit.

\begin{figure}[htbp]
 \includegraphics[width=0.6\linewidth]{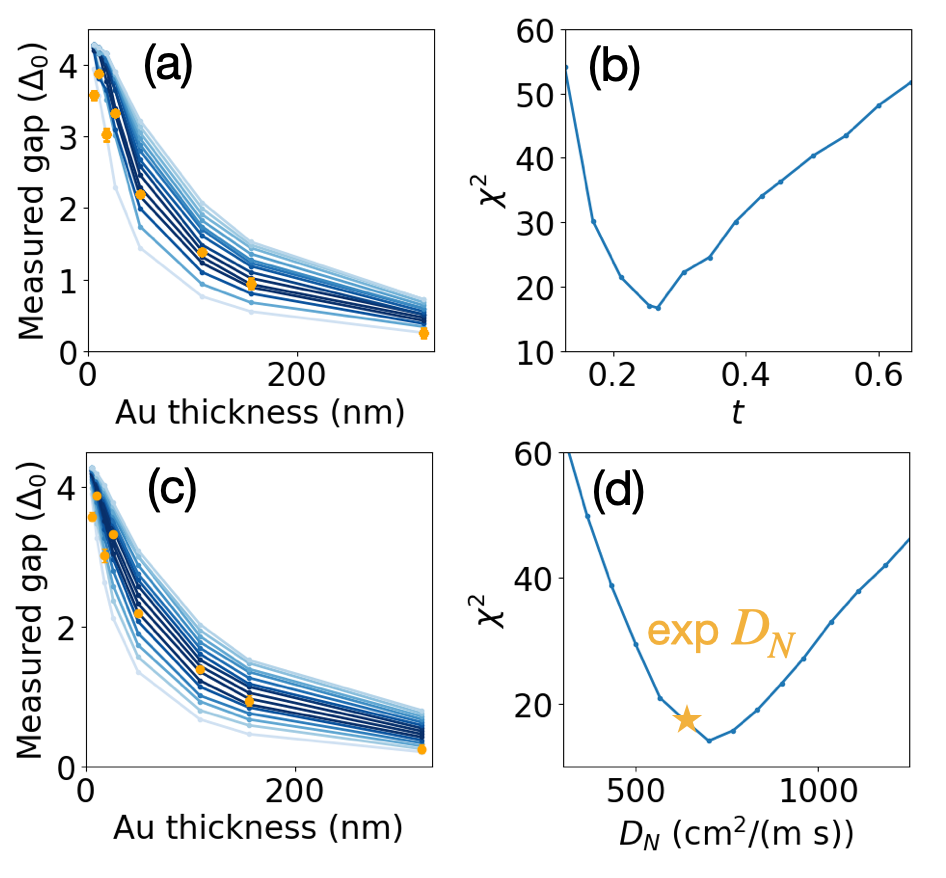}
\caption{\label{fig:usadel_confidence} Model curves of minigap dependence on Au thickness when sweeping (a) \(t\) and (c) \(D_N\) near the optimal parameters. \(t\) is swept from 0.13 to 0.65 and \(D_N\) is swept from 300 to 1250 cm$^2$/(m$\cdot$s). (b) and (d) display the $\chi^2$ values for the family of curves in (a,b). The transparency of 0.24 is fit through minimizing $\chi^2$ in (b). (d) confirms that our experimental extraction of $D_N$ (marked with star) is accurate.}
\end{figure}

To estimate an uncertainty range for the fitted transparency value, we sweep \(t\) around the optimal fitted value and plot \(\chi^2\) values at different \(t\) values in Fig. \ref{fig:usadel_confidence}. Note that the \(\chi^2\) values are relatively large since we are underestimating the true statistical fluctuations in the superconducting gap. Since we fit \(t\) by optimizing the \(\chi^2\) value, the minimum is exactly the value of \(t=0.24\) as reported in the main text. The uncertainty range in the main text is derived from the values of $t$ where the value of $\chi^2$ reaches twice the minimal value. Similarly, sweeping \(D_N\) results in a minimum at 700 cm$^2$/(m$\cdot$s), which is within the error range of the experimental estimation of 622 cm$^2$/(m$\cdot$s).

\clearpage


\bibliography{truncated_bib}